%% file: main.tex
\documentclass[%
 reprint, 
 superscriptaddress,
 amsmath,amssymb,
 aps,
floatfix,
longbibliography
]{revtex4-1}

\usepackage{graphicx}
\usepackage{bm}

\usepackage{xcolor}
\usepackage[colorlinks=true,citecolor=blue,linkcolor=magenta]{hyperref}

\newcommand{\la}{\langle}
\newcommand{\ra}{\rangle}
\newcommand{\rar}{\rightarrow}
\newcommand{\da}{\downarrow}
\newcommand{\ua}{\uparrow}
\newcommand{\ben}{\begin{eqnarray}}
\newcommand{\een}{\end{eqnarray}}
\newcommand{\be}{\begin{equation}}
\newcommand{\ee}{\end{equation}}

\begin{document}

\title{Nonadiabatic Phase Transition with Broken Chiral Symmetry}

\author{Bin Yan}
\affiliation{Center for Nonlinear Studies, Los Alamos National Laboratory, Los Alamos, New Mexico 87545, USA}
\affiliation{Theoretical Division, Los Alamos National Laboratory, Los Alamos, New Mexico 87545, USA}

\author{Vladimir Y. Chernyak}
\affiliation{Department of Chemistry and Department of Mathematics, Wayne State University, Detroit, Michigan 48202, USA}

\author{Wojciech H. Zurek}
\affiliation{Theoretical Division, Los Alamos National Laboratory, Los Alamos, New Mexico 87545, USA}

\author{Nikolai A. Sinitsyn}
\affiliation{Theoretical Division, Los Alamos National Laboratory, Los Alamos, New Mexico 87545, USA}


\begin{abstract}
We explore nonadiabatic quantum phase transitions in  an Ising spin chain with a linearly time-dependent transverse field and two different spins per unit cell. Such a spin system passes through critical points with gapless excitations, which support nonadiabatic transitions. 
Nevertheless, we find that the  excitations on one of the chain sublattices are suppressed in the nearly adiabatic regime exponentially. Thus, we reveal a coherent mechanism to induce exponentially large  density separation for different quasiparticles. 
\end{abstract}

\maketitle

The task to separate two slightly different types of particles is often encountered in both applied and fundamental research. One example is the apparent asymmetry between matter and antimatter in our Universe. Observations prove that the symmetry between matter and antimatter was broken, presumably early on in the history of the Universe. However, it is still unclear how the subtle $CP$ symmetry violation could lead to the large observed differences at the cosmological scale \cite{cp-sym}, although it is known that, when the characteristic asymmetry energy scale is very small, the large density difference has to be produced during a nonequilibrium process \cite{Sakharov}. 

The goal of this Letter is to introduce a mechanism for controlling different quasiparticles separately using exponential sensitivity of quantum nonadiabatic transitions to  symmetry-breaking interactions. 
Namely, when parameters of a quantum system change with time adiabatically, the system remains in the instantaneous eigenstate, e.g., the ground state. However, the Kibble-Zurek mechanism~\cite{Kibble1980-to,Kibble2001-so,Kibble2007-rz,Zurek1985-cs,Zurek1996-xm,Del_Campo2014-rj,KZ-exp,Rams2019-kt,Qiu-exp} predicts that the number of nonadiabatic excitations is not suppressed exponentially if a macroscopic system passes through a quantum critical point at which the energy gap to excitations closes. Without an asymmetry of interactions, different particle species would pass through the same  phase transition simultaneously. However, even a small asymmetry generally opens the gap to certain excitation types, even though the critical point is not destroyed. Thus, we can harvest the excitations of one type and suppress the others.

\begin{figure}[t!]
    \centering
    \includegraphics[width=\columnwidth]{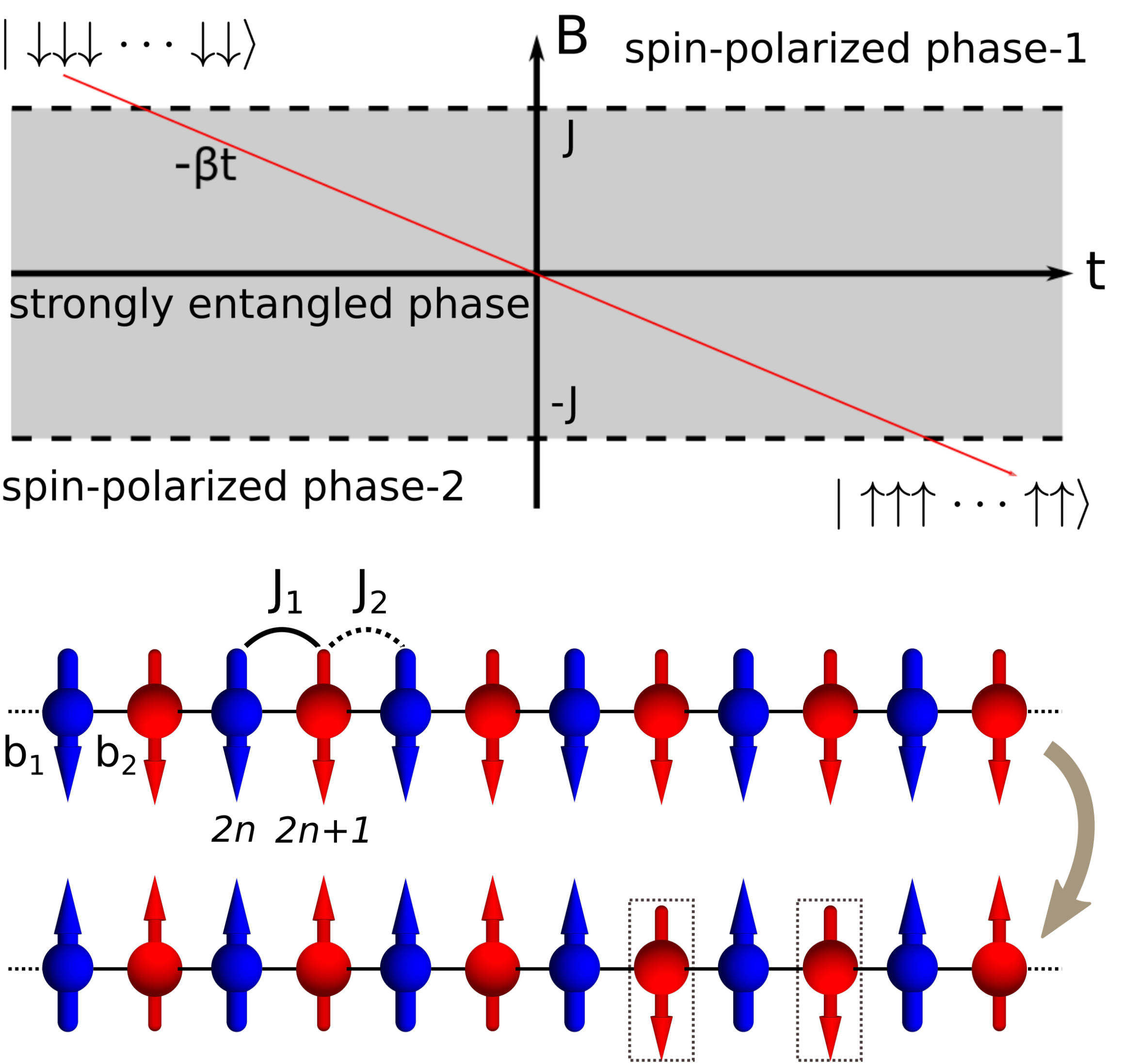}
    \caption{Top: phase diagram of Ising chain (\ref{H1-chain}) in a transverse magnetic field. Dashed horizontal lines mark the critical points with the spectral gap closings. At $B=\pm \infty$, the ground state is fully spin polarized along the $z$ axis, whereas at $B=0$ the ground state is a superposition of $x$-polarized spin states. Bottom: spin chain model (\ref{Ham}) with broken chiral symmetry. We consider the case of $b_1>b_2$. In the adiabatic limit, the initial spin polarized ground state (top) at $t=-\infty$ transfers to the spin polarized state at $t=+\infty$ with all spins flipped. The nonadiabatic excitations (nonflipped spins) on the odd-site sublattice (red) are suppressed as a power law (\ref{scale3}), respecting the Kibble-Zurek mechanism. In contrast, nonadiabatic excitations on the even-site sublattice (blue) are suppressed exponentially (\ref{scale1}).}
    \label{fig:Ising-phase}
\end{figure}

A broadly known quantum example that confirms the Kibble-Zurek mechanism is the model of a uniform Ising spin chain in a transverse magnetic field~\cite{Dziarmaga2005,Zurek2005-mn,Heyl2013-vw}, with the Hamiltonian
\begin{equation}\label{H1-chain}
        \hat{H}_{u} = \sum_{n=1}^N \left[ B \hat{\sigma}_{n}^z + 
        J\hat{\sigma}_{n}^x\hat{\sigma}_{n+1}^x \right], \quad B=-\beta t,
\end{equation}
where $\hat{\sigma}_n^{x,z}$ are Pauli operators for the $n$th spin, $J$ is the spin-spin coupling, and $B$ is the transverse time-dependent field that changes with rate $\beta > 0$. This model has two Dirac points at $B=\pm J$, at which its spectral gaps close and which mark boundaries between three phases. The phase with strong quantum correlations (Fig.~\ref{fig:Ising-phase}, top) contains the point $B=0$ with two degenerate ground states:
\be
|\rar\,\rar \, \cdots \rar\,\rar \ra \quad {\rm and}\quad |\leftarrow\,\leftarrow\, \cdots \leftarrow\,\leftarrow \ra.
\label{gs2}
\ee

For  $B(t)=-\beta t$, the model is exactly solvable~\cite{Dziarmaga2005}. If the system starts from the fully polarized ground state
\begin{equation}\label{gs1}
|\da\,\da \,\da\,\da \ldots \da\,\da \ra
\end{equation}
as $t\rar -\infty$, then
in the thermodynamic limit of this solution, after the passage through a critical point, the number of nonadiabatic excitations follows a power law:
\begin{equation}\label{n-kinks}
\rho \sim J^{-1}\beta^{1/2}.
\end{equation}

Let us now explore robustness of the Ising model predictions by considering a spin chain whose unit cell contains two different spins, with the Hamiltonian
\begin{equation}\label{Ham}
\begin{aligned}
        \hat{H} =& \sum_{n=1}^N \left[ -\frac{b_1t}{2}\hat{\sigma}_{2n}^z - \frac{b_2t}{2}\hat{\sigma}_{2n+1}^z \right] \\
        &+\sum_{n=1}^N\left[ J_1\hat{\sigma}_{2n}^x\hat{\sigma}_{2n+1}^x + J_2\hat{\sigma}_{2n+1}^x\hat{\sigma}_{2n+2}^x \right].
\end{aligned}
\end{equation}
The difference between $b_1$ and $b_2$ is due to different spin $g$ factors, and $J_1\ne J_2$ reflects the lack of inversion symmetry at zero field. A periodic boundary condition is imposed. Figure~\ref{fig:Ising-phase} (bottom) illustrates the structure of this spin chain.

Without loss of generality, here and in what follows, we assume $b_1>b_2$; namely, the spins on the even-site sublattice have a larger $g$ factor than those on the odd sites.
We note that, for weak symmetry breaking, $|J_1-J_2|\ll |J_1+J_2|$ and $|b_1-b_2|\ll |b_1+b_2|$, the ground state polarizations at $B=0,\pm\infty$ are the same for spins with odd and even indices, and the spin excitations on different sublattices then resemble particles of two slightly different types of matter.

Via the Jordan-Wigner  transformation, Hamiltonian (\ref{Ham}) reduces to a quadratic form
of fermionic operators, $\hat{c}$ and $\hat{d}$, on the even and odd sites, respectively. The details  are presented in Appendix \ref{app:JW} of Supplemental Material~(SM). Thus, 
\begin{equation}
\hat{\sigma}_{2n}^z = 1 - 2 \hat{c}^\dag_n\hat{c}_n, \ \ \ \hat{\sigma}_{2n+1}^z = 1 - 2 \hat{d}^\dag_n\hat{d}_n.
\end{equation}
It is convenient to work with the Fourier transformed operators
\begin{equation}
    \hat{c}_p = \frac{1}{\sqrt{N}}\sum_{n=1}^N \hat{c}_n e^{-ipn}, \ \ \  \hat{c}^\dag_p = \frac{1}{\sqrt{N}}\sum_{n=1}^N \hat{c}^\dag_n e^{ipn},
\end{equation}
and similarly defined $\hat{d}_p$ and $\hat{d}^\dag_p$. 
We will assume that  $N$ is even, so the momentum takes discrete values $p=\pm(2k-1)\pi/N$, $k=1,\ldots,N/2$. Hamiltonian (\ref{Ham}) then is
\begin{equation}
    \hat{H} = \sum_{p>0} \hat{\bm{a}}^\dag_p {H}_p \hat{\bm{a}}_p,
\end{equation}
where $\hat{\bm{a}}_p$ and ${H}_p$ are given by, respectively,
\begin{equation}\label{ham-4}
\hat{\bm{a}}_p \equiv
\begin{pmatrix}
\hat{c}_p\\
\hat{c}^\dag_{-p}\\
\hat{d}_p\\
\hat{d}^\dag_{-p}\\
\end{pmatrix},
 \ \ \ H_p \equiv
\begin{pmatrix} 
b_1  t     & 0            & g     & \gamma  \\
0         & -b_1 t         & -\gamma    &-g  \\
  g^*  &-\gamma^*        & b_2t  & 0       \\
 \gamma^* & -g^*    & 0     & -b_2 t     \\ 
\end{pmatrix},
\end{equation}
with the couplings 
\begin{equation}
\label{couplings}
g \equiv J_1 + J_2 e^{-ip}, \ \ \gamma \equiv J_1 - J_2 e^{-ip}.
\end{equation}
In the Heisenberg picture, the evolution of $\hat{\bm{a}}_p$ is described by a Schr\"{o}dinger-like equation:
\begin{equation}\label{eq:evolution}
i \hat{\bm{a}}_p(t) = H_p(t) \hat{\bm{a}}_p (t),
\end{equation}
and the initial ground state (\ref{gs1}) corresponds to the initially fully filled Fermi sea, for all fermions as $t\rar -\infty$. 

\begin{figure}[t!]
    \centering
    \includegraphics[width=0.95\columnwidth]{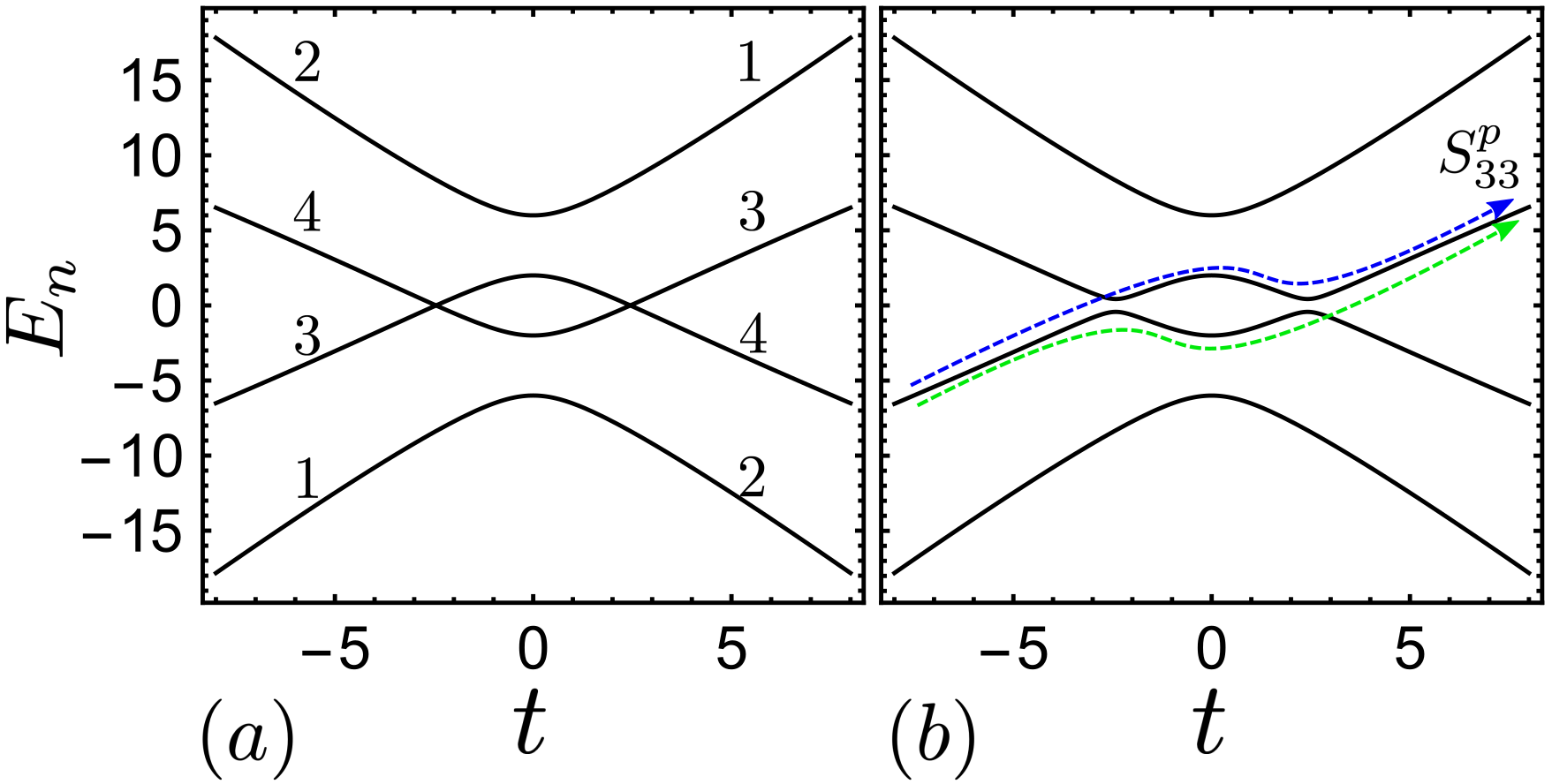}
    \caption{The time-dependent spectrum of the Hamiltonian $H_p(t)$ from Eq.~(\ref{ham-4}). (a) Two exact level crossing points appear at $p=0$. (b) At finite $p$, which is here $p=0.3$, small gaps open up near places of the former crossing points. Other parameters are $b_1=2$, $b_2=1$, $J_1=3$, and $J_2=1$ in both figures. The numbers $1$-$4$ in (a) mark the diagonal elements of $H_p$, with which the corresponding eigenenergies merge as $t\rar \pm \infty$, e.g., $1$ is for $(H_p)_{11}$. The dashed blue and green arrows in (b) show two semiclassical trajectories that contribute to the amplitude $S_{33}$. Each trajectory jumps through one of the gaps but avoids the other gap.   }
    \label{fig:spectrum4}
\end{figure} 

In Fig.~\ref{fig:spectrum4}, we show the time-dependent eigenvalues of the matrix $H_p$ for a broken chiral  symmetry: $J_1\ne J_2$ and $b_1> b_2$. Figure~\ref{fig:spectrum4}(a) shows that at $p=0$ two of the four energy levels experience exact level crossings, whereas Fig.~\ref{fig:spectrum4}(b) illustrates that this degeneracy is lifted for  nonzero $p$. This means that the chiral asymmetry  does not destroy the Dirac points in the spectrum, which must produce a power-law density of excitations according to the Kibble-Zurek mechanism. 

We will consider the evolution with the Hamiltonian (\ref{Ham}) during the time interval $t\in (-\infty,+\infty)$. In terms of the phase diagram in Fig.~\ref{fig:Ising-phase}, this means that the system passes through two phase transition points. 
In the adiabatic limit, the initial ground state (\ref{gs1}) transfers, as $t\rar +\infty$, into the ground state with all spins up
\begin{equation}\label{gs3}
|\ua\,\ua \,\ua\,\ua \ldots \ua\,\ua \ra.
\end{equation}
This is the fermionic vacuum, and the free fermions are the elementary excitations. The numbers of  excitations (nonflipped spins) on the even and odd sublattices  at the end of the protocol are given by, respectively, 
\ben
\label{n1-def}
N_{\rm ex}^e\equiv \sum_{n=1}^N \la \hat{c}_n^{\dagger} \hat{c}_n \ra = \sum_{p}^N \la \hat{c}_p^{\dagger} \hat{c}_p \ra,\\
\label{n2-def}
N_{\rm ex}^o\equiv \sum_{n=1}^N \la \hat{d}_n^{\dagger} \hat{d}_n \ra =\sum_{p}^N \la \hat{d}_p^{\dagger} \hat{d}_p \ra,
\een
where, in the Schr\"odinger picture, the averaging is taken over the state at $t=+\infty$.

The Hamiltonian (\ref{ham-4}) has the form of a  multistate Landau-Zener (MLZ) model, i.e.,  can be written as $H(t)=A+Bt$ with Hermitian matrices $A$ and $B$. The MLZ models have been extensively studied previously~\cite{be-lz,commute,four-LZ}. Our case with $J_1\ne J_2$ is generally not solvable \cite{nonsol-lz} but  the MLZ theory provides exact formulas for some of the evolution matrix elements \cite{onepoint-LZ}, which we will utilize.

Let us define the $S$ matrix
\begin{equation}
S=S(p) \equiv  U^p (T,-T)_{T\rar \infty},
\label{s-def}
\end{equation}
where $U^p(T,-T)$ is the evolution matrix over the time interval $t\in(-T,T)$ with the Hamiltonian ${H}_p$.
We can  say that the operators at $t=\pm \infty$ are related by
\begin{equation}
    \hat{\bm{a}}_p(+\infty) = S\hat{\bm{a}}_p,
\end{equation}
where $\hat{\bm{a}}_p \equiv \hat{\bm{a}}_p (-\infty)$. Hence, 
\begin{equation}
\begin{aligned}
     &   \hat{c}_p(+\infty) = S_{11}\hat{c}_p + S_{12}\hat{c}^\dag_{-p} + S_{13}\hat{d}_p + S_{14}\hat{d}^\dag_{-p},\\
&    \hat{c}^\dag_p(+\infty) = S^*_{11}\hat{c}^\dag_p + S^*_{12}\hat{c}_{-p} + S^*_{13}\hat{d}^\dag_p + S^*_{14}\hat{d}_{-p}.
\end{aligned}
\end{equation}
The number of excitations~(\ref{n1-def}) can be evaluated in the Heisenberg picture, i.e., taking the average of the  operators at $t=+\infty$ with respect to the initial state~(\ref{gs1}):
\begin{equation}
\label{mag1}
    N_{\rm ex}^e(+\infty) = \sum_p \ (|S_{11}|^2 + |S_{13}|^2).
\end{equation}
The survival amplitudes for states with the highest-energy level slopes are known exactly for any MLZ  model~\cite{be-lz}:
\begin{equation}
\label{hyer1}
    S_{11}=S_{22}= e^{ -\pi|g|^2/(b_1-b_2) -\pi |\gamma|^2/(b_1+b_2)}.
\end{equation}
Another exact relation of MLZ theory is for the transition amplitudes between the levels with the two highest  slopes~\cite{onepoint-LZ}:
\be
S_{11}S_{33}+|S_{13}|^2 = e^{  -2\pi |\gamma|^2/(b_1+b_2)}.
\label{hyer2}
\ee
This does not fix $S_{13}$ because $S_{33}$ is not known. Fortunately, 
model (\ref{ham-4}) has discrete symmetries leading to
\begin{equation}
\label{disc}  
\begin{aligned}
   &S_{11}+S_{22}=S_{33}+S_{44},\\
 & S_{33}(-p) = S_{44}(p),
\end{aligned}
\end{equation}
the derivations of which are given in Appendix \ref{app:s13} of SM.
Equations~(\ref{hyer1})-(\ref{disc}) then predict
\begin{equation}\label{eq:s13}
     \sum_{p} |S_{13}|^2 = \sum_{p}\left[e^{-2\pi |\gamma|^2/(b_1+b_2)}- |S_{11}|^2 \right].
\end{equation}
This formula is exact. Hence, without approximations
\begin{equation}
    N^e_{\rm ex}(+\infty) = \sum_p\  e^{-2\pi|\gamma|^2/(b_1+b_2)}.
\label{fin1}
\end{equation}

\begin{figure}[t!]
    \centering
    \includegraphics[width=\columnwidth]{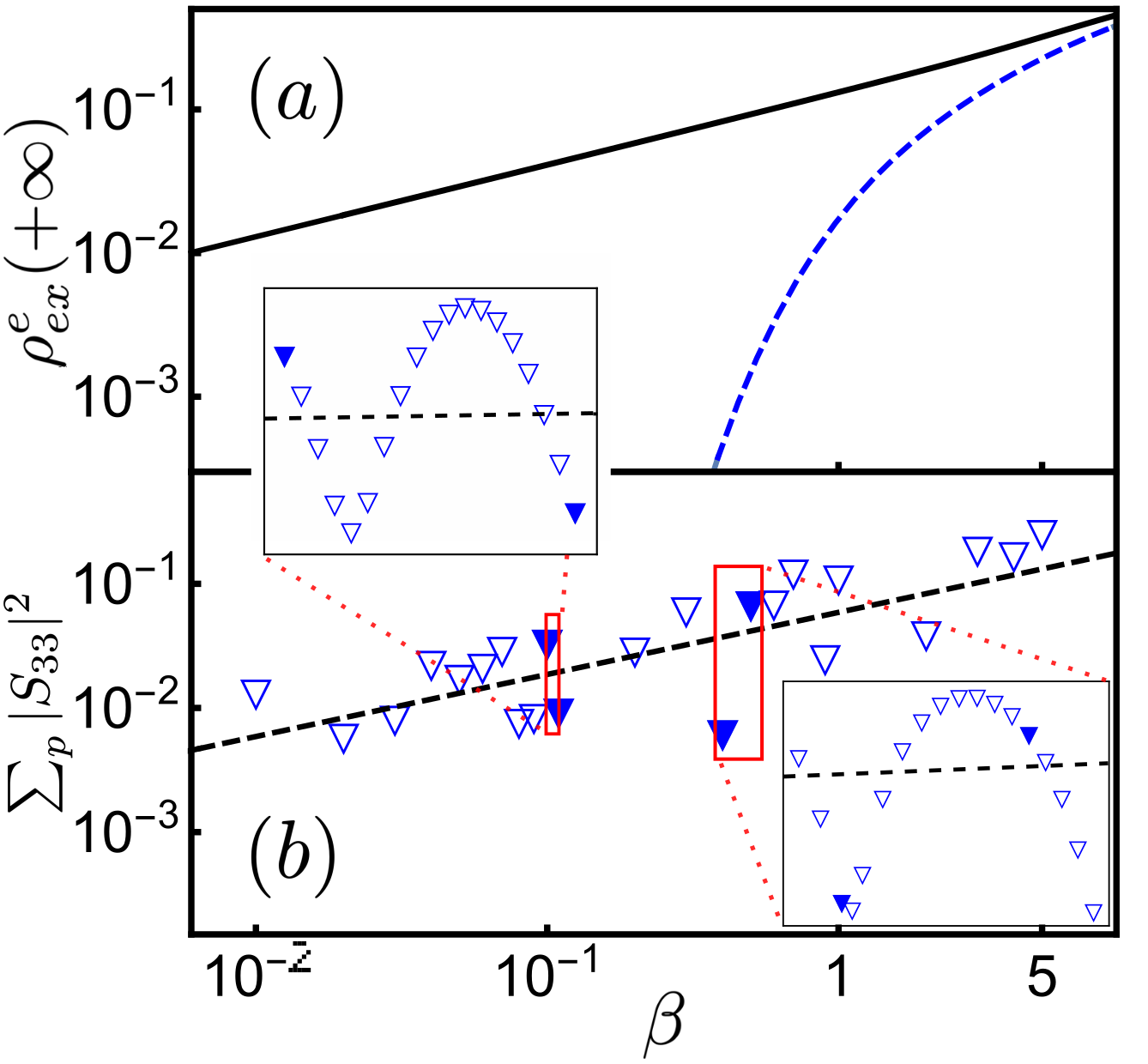}
    \caption{(a) Density of excitations of the even-site sublattice as a function of the quenching rate, $b_1=2\beta$, $b_2=\beta$. The black solid curve and blue dashed curve correspond to, respectively, the exact prediction~(\ref{eq:mexact}) at $J_1=J_2=1.5$ and perturbed coupling $J_1=2$, $J_2=1$. (b) The numerically calculated  transition probability $\sum_p|S_{33}|^2$ at $J_1=2$, $J_2=1$, $b_1=2\beta$, and $b_2=\beta$. Triangles are the numerical data. The black dashed line is the power-law scaling~(\ref{scale3}) prediction. The insets resolve the  fast Stueckelberg oscillations around the mean power-law tail. The filled triangles are the same points in the main figure and the insets.}
    \label{fig:excitation}
\end{figure} 

As $N\rightarrow\infty$, we replace $\sum_p/N \rar\int_{-\pi}^{\pi}dp/(2\pi)$ and find the
density of excitations (per unit cell) in the thermodynamic limit
\begin{equation}\label{eq:mexact}
\rho^e_{\rm ex}(+\infty) = \exp\left(-\frac{2\pi (J_1^2+J_2^2)}{|b_1+b_2|}\right) {I}_0\left(\frac{4\pi J_1J_2}{b_1+b_2}\right),
\end{equation}
where ${I}_0(x)$ is the modified Bessel function of the first kind. In the leading order
of small $b_1+b_2$ 
\begin{equation}
\label{scale1}
    \rho^e_{\rm ex}(+\infty) \approx \exp{\left(-\frac{2\pi (J_1-J_2)^2}{|b_1+b_2|}\right)}\frac{\sqrt{b_1+b_2}}{2\pi\sqrt{2J_1J_2}}.
\end{equation}
This is the central result of our Letter. For any $J_1\ne J_2$, the excitations on the even sublattice are  suppressed exponentially rather than by a power law. The latter is found  only for  a symmetric case with $J_1=J_2$ but $b_1>b_1$. Complete exact solution for this special case is presented in Appendix \ref{app:exact} of SM. We also note that the above solutions do not smoothly carry over the $b_1=b_2$ point, to which our exact MLZ analysis does not apply because of the degeneracy of the time-dependent Zeeman coupling.

Equation~(\ref{scale1}) does not contradict the Kibble-Zurek mechanism, because the power law for excitation density does appear on the odd-site sublattice. Consider
\begin{equation} \label{n2}
     N^o_{\rm ex}(+\infty) = \sum_p (|S_{33}|^2 +|S_{31}|^2).
\end{equation}
Here, $|S_{31}|^2$ can be obtained exactly, because $|S_{31}|=|S_{13}|$. According to Eq.~(\ref{eq:s13}), it produces an exponentially suppressed contribution. We also prove in Appendix \ref{app:bounds} of SM that $\sum_p |S_{33}|^2\ge \sum_{p} |S_{11}|^2$, so $N^o_{\rm ex} \ge N^e_{\rm ex}$. Hence, only $S_{33}$ contains the information about the power-law scaling.
 
The exact crossings of the second and third adiabatic levels of $H_p$ happen at $p=0$ and time moments
\begin{equation}\label{tpt}
t_{1,2}=\mp 2\sqrt{J_1J_2/(b_1b_2)}.
\end{equation}
We interpret $t_{1,2}$ as the moments of the phase transitions.  By setting $t=t_{1,2}+\delta t$ and $p=\delta p$, with small $\delta p$ and $\delta t$, we find an effective Hamiltonian for the interactions within the subspace of $H_p$ spanned by the two eigenstates, whose energies become nearly degenerate near the critical points:
\begin{eqnarray}
\label{gap}
\hat{H}_{\rm eff} \approx \frac{2\sqrt{J_1J_2b_1b_2}( \sqrt{J_1 J_2} \delta p \hat{\tau}_x + \sqrt{b_1b_2} \delta t \hat{\tau_z} )}{\sqrt{(b_2J_1+b_1J_2)(b_1J_1+b_2J_2)}},
\end{eqnarray}
where $\hat{\tau}_x$ and $\hat{\tau}_z$ are the Pauli operators that act in subspace of two states with closest   to one Dirac point energies.
The Landau-Zener formula applied to  Hamiltonian~(\ref{gap}) returns the probability of a nonadiabatic excitation after a transition through one Dirac point:
\begin{equation}\label{Pna}
P_{\rm ex}=\exp \left({- \frac{ 2\pi (\delta p)^2(J_1J_2)^{3/2}}{\sqrt{(b_2J_1+b_1J_2)(b_1J_1+b_2J_2)}}}\right). 
\end{equation}

Our system passes through two Dirac points. An excitation created near the first point should remain excited after  the second  point. The system may also not produce an excitation during the first crossing but produce it at the second one. Generally, there is interference between such evolution paths (Stueckelberg's oscillations). If we disregard this interference,  the probabilities of the two possibilities  simply sum, i.e.,
\begin{equation}\label{pexfin}
P_{\rm 2ex}=2P_{\rm ex}(1-P_{\rm ex}).
\end{equation}
Integrating $P_{\rm 2ex}$ over $\delta p \in (-\infty, +\infty)$, we find that in the adiabatic limit
\begin{equation}\label{scale3}
\sum_p\frac{|S_{33}|^2}{N} \approx \frac{2-\sqrt{2}}{4\pi(J_1J_2)^{3/4}} \left[(b_2J_1+b_1J_2)(b_1J_1+b_2J_2) \right]^{1/4}.
\end{equation}
This is the estimate of the excitation density on the odd-site sublattice in the nearly adiabatic limit. In Fig.~\ref{fig:excitation}, the numerical simulations confirm the power-law scaling trend for the parameter dependence in Eq.~(\ref{scale3}), which is modulated by fast Stueckelberg oscillations. We also note that Eq.~(\ref{scale3}) reproduces correctly  the scaling prediction of the uniform chain model~(\ref{n-kinks}) if we set $b_1=b_2=2\beta$ and $J_1=J_2$.

Hence, as $t\rightarrow+\infty$, the remaining excitations on the odd-site sublattice are suppressed according to a power law, in contrast to the exponential suppression on the even-site sublattice. At the intermediate moment $t= 0$, excitations take the form of superpositions of kinks. The asymmetry appears then, too, i.e., some of the excitations are exponentially suppressed in comparison to the others but this effect does not have a simple interpretation in terms of different kink types. We leave such details to Appendix \ref{app:kinks} of SM.

Our results illustrate  how the Kibble-Zurek mechanism works when  a part of a system is not observable. The latter may happen in atomic gases, in which optically observable spins interact via spins of other atomic species \cite{twobeam}. In the asymmetric Ising chain, if the spin excitations on the odd sites are not observable, the even-site spins still create a visible ferromagnetic state at zero external field and polarize when the external field is strong. Hence, looking only at the even spins, one can expect that this system goes through the same phase transitions as the uniform Ising chain and, thus, produces excitations with a power-law scaling. Our solution shows, however, that all observable spin excitations are  suppressed in this case exponentially. A power-law tail of the excitations is hidden then in the nonobservable spins.

In summary, we demonstrated that small differences between interacting spins in a simple spin chain can lead to an exponentially large effect when passing slowly through a phase transition. 
The underlying mechanism arguably processes universalities: Strong but symmetric interactions compete with each other near a critical point, where subtle interaction differences  play a decisive role. As long as a symmetry breaking opens the gap for certain types of quasiparticles while not  destroying the critical point, the exponentially large density separation of different quasiparticles should happen after the passage through the phase transition independently of the model's microscopic details.

The Kibble-Zurek mechanism for Ising-type quantum phase transition has been recently studied experimentally in ultracold atoms~\cite{KZ-exp}. In such systems, an asymmetry can be introduced by placing atoms in a periodic potential without the inversion symmetry. If the time-linear field ramp is induced by changing the ac frequency in the rotating-wave approximation, the sweeping of the detuning frequency across an optical resonance effectively mimics the field changes in range $B_z\in (+\infty,-\infty)$ \cite{kramers-sun,adiab-fr}. Hence, the demonstration of our effect requires only simple modification of the control protocols in already accessible for studies dynamic phase transitions. 

Let us also speculate about possible applications. Different isotopes have different spin interactions in an ultracold atomic mixture \cite{isotopes}. Hence, an adiabatic passage through a quantum critical point can   induce spin excitations in this mixture, overwhelmingly, for only one of the isotopes. One can then separate such excited atoms using the magnetic deflection approach from Ref.~\cite{isotope-exp}, and, thus, develop a technology for isotope separation.  The asymmetry of matter and antimatter in our Universe may also be viewed now as a result of a hypothetical transition through a quantum critical point during cosmic inflation, when the matter could be excited from the vacuum via quantum nonadiabatic processes \cite{inflation,Enomoto2020-ji}.

The dynamic phase transitions are found broadly, from  cosmological models to the experiments with superconductors and ultracold atoms. 
They are likely responsible for the scaling of mistakes that limit the quantum annealing computation techniques \cite{q-annealing}.
Fortunately, different systems show universal behavior, which is driven by relatively simple effects near the critical point. The quasiparticle separation is one of such effects that remain relevant near the critical point and, thus, can be used for control of complex quantum systems. 
  
\begin{acknowledgements}
This work was carried out under the support of the U.S. Department of Energy, Office of Science, Basic Energy Sciences, Materials Sciences and Engineering Division, Condensed Matter Theory Program. B.Y. also acknowledges partial support from the Center for Nonlinear Studies.
\end{acknowledgements}

\bibliography{reference}

\clearpage
\input{supp}

\end{document}

%% file: supp.tex

\appendix

\setcounter{page}{1}
\renewcommand\thefigure{\thesection\arabic{figure}}
\setcounter{figure}{0} 

\onecolumngrid

\begin{center}
\large{ Supplemental Material for \\ ``Nonadiabatic Phase Transition with Broken Chiral Symmetry''
}
\end{center}

\section{Jordan-Wigner transformation}\label{app:JW}

For the model Hamiltonian
\begin{equation}\label{supp:Ham1}
        \hat{H} = \sum_{n=1}^N \left[ -\frac{b_1t}{2}\hat{\sigma}_{2n}^z - \frac{b_2t}{2}\hat{\sigma}_{2n+1}^z \right] +\sum_{n=1}^N\left[ J_1\hat{\sigma}_{2n}^x\hat{\sigma}_{2n+1}^x + J_2\hat{\sigma}_{2n+1}^x\hat{\sigma}_{2n+2}^x \right]
\end{equation}
with periodic boundary condition, we define two seperate sets of Jordan-Wigner transformation for the sublattices with even and odd indices, i.e.,
\begin{equation}\label{app:JWT}
\begin{aligned}
 &\hat{\sigma}_{2n}^z = 1 - 2 \hat{c}^\dag_n\hat{c}_n, \\  &\hat{\sigma}_{2n+1}^z = 1 - 2 \hat{d}^\dag_n\hat{d}_n,\\
 &\hat{\sigma}_{2n}^x = -(\hat{c}_n^\dag+\hat{c}_n)\prod_{m<n}(1 - 2 \hat{c}^\dag_m\hat{c}_m)(1 - 2 \hat{d}^\dag_m\hat{d}_m),\\
 &\hat{\sigma}_{2n+1}^x = -(\hat{d}_n^\dag+\hat{d}_n)(1-2\hat{c}_n^\dag \hat{c}_n)\prod_{m<n}(1 - 2 \hat{c}^\dag_m\hat{c}_m)(1 - 2 \hat{d}^\dag_m\hat{d}_m),
\end{aligned}
\end{equation}
where 
\begin{equation*}
\begin{aligned}
&\{\hat{c}_n,\hat{c}_{n'}\}=\{\hat{c}^\dag_n,\hat{c}^\dag_{n'}\}=\{\hat{d}_n,\hat{d}_{n'}\}=\{\hat{d}^\dag_n,\hat{d}^\dag_{n'}\}=0,\\
&\{\hat{c}_n,\hat{d}_{n'}\}=\{\hat{c}^\dag_n,\hat{d}^\dag_{n'}\}=\{\hat{c}_n,\hat{d}^\dag_{n'}\}=\{\hat{c}^\dag_n,\hat{d}_{n'}\}=0,\\
& \{\hat{c}_n,\hat{c}^\dag_{n'}\}=\{\hat{d}_n,\hat{d}^\dag_{n'}\}=\delta_{n,n'}.\\
\end{aligned}
\end{equation*}
Their Fourier transforms are defined as
\begin{equation}\label{app:FT}
    \hat{c}_p = \frac{1}{\sqrt{N}}\sum_{n=1}^N \hat{c}_n e^{-ipn}, \ \ \  \hat{c}^\dag_p = \frac{1}{\sqrt{N}}\sum_{n=1}^N \hat{c}^\dag_n e^{ipn}.
\end{equation}
Due to the periodic boundary condition of the spin chain, the discrete momentum takes values in the range $(-\pi,\pi)$,
\begin{equation}
\begin{aligned}
        &p=\pm(2k-1)\pi/N,\ \  k=1,\ldots,N/2,\ \ N \text{ even};\\
        &p=\pm 2k\pi/N,\ \  k=1,\ldots,(N-1)/2,\ \ N \text{ odd}.
\end{aligned}
\end{equation}
In terms of the fermionic operators, the original spin chain Hamiltonian (\ref{supp:Ham1}) transforms to 
\begin{equation}
    \hat{H} = \sum_{p>0} \hat{\bm{a}}^\dag_p H_p \hat{\bm{a}}_p,
\end{equation}
where $\hat{\bm{a}}_p$ and $H_p$ are defined as
\begin{equation}\label{supp:Ham4}
\hat{\bm{a}}_p \equiv
\begin{pmatrix}
\hat{c}_p\\
\hat{c}^\dag_{-p}\\
\hat{d}_p\\
\hat{d}^\dag_{-p}\\
\end{pmatrix},
 \ \ \ H_p \equiv
\begin{pmatrix} 
b_1  t     & 0            & g     & \gamma  \\
0         & -b_1 t         & -\gamma    &-g  \\
  g^*  &-\gamma^*        & b_2t  & 0       \\
 \gamma^* & -g^*    & 0     & -b_2 t     \\ 
\end{pmatrix},
\end{equation}
with the couplings 
\begin{equation}
\label{couplings}
g \equiv J_1 + J_2 e^{-ip}, \ \ \gamma \equiv J_1 - J_2 e^{-ip}.
\end{equation}

\section{Exact solution for $\sum_p |S_{13}|^2$}\label{app:s13}

Consider the $S$ matrix in the 4-level Landau-Zener problem of the Hamiltonian $H_p$, which is defined in Eq.~(\ref{supp:Ham4}). Here, the $S$ matrix refers to the unitary matrix \begin{equation}
S = S(p) \equiv  U^p (T,-T)_{T\rar \infty},
\label{s-def}
\end{equation}
where $U^p(T,-T)$ is the evolution matrix over the time interval $t\in(-T,T)$ with the Hamiltonian $H_p$.

This model belongs to a particularly well-understood  subclass of multilevel Landau-Zener (MLZ) problems. Namely, it contains bipartite interactions and all diabatic levels are crossing in one point in the time-energy diagram \cite{onepoint-LZ}. Here ``bipartite" means that
all states can be partitioned into two groups, such that states in one group are coupled directly only to states in the other group; and the diabatic levels are linearly time-dependent diagonal elements of the Hamiltonian.  Following Ref.~\cite{onepoint-LZ}, we list here three properties of the $S$ matrix that are essential for solving the exact solutions:

(i) As in all other MLZ systems, the   amplitudes of surviving on the levels with the lowest and highest slopes are given by the Brundobler-Elser formula:
\begin{subequations}\label{be}
\be
\label{be1}
S_{11}=S_{22} = e^{ -\frac{\pi|g|^2}{b_1-b_2} -\frac{\pi |\gamma|^2}{b_1+b_2}}.
\ee
Considering the definition of $g$ and $\gamma$ in (\ref{couplings}), this means also that 
\be
S_{11}(p)=S_{22}(p) = S_{11}(-p)=S_{22}(-p).
\label{be2}
\ee
\end{subequations}

(ii) The other two diagonal elements of the scattering matrix, $S_{33}$ and $S_{44}$, can be different from each other and from Eq.~(\ref{be1}), but they are real and satisfying the relations (see Eq.~(21) in Ref.~\cite{onepoint-LZ}):
\begin{subequations}
\begin{equation}\label{one-rel1}
S_{11}S_{33}+|S_{13}|^2=e^{  -\frac{2\pi |\gamma|^2}{b_1+b_2}}
\end{equation}
\begin{equation}\label{one-rel2}
S_{44}S_{22}+|S_{42}|^2=e^{  -\frac{2\pi |\gamma|^2}{b_1+b_2}}
\end{equation}
\end{subequations}

(iii) Following Ref.~\cite{onepoint-LZ}, we can derive one more constraint for the diagonal elements of the $S$ matrix. Define a matrix
\begin{equation*}
    \Theta \equiv
\begin{pmatrix} 
\hat{\mathbb{I}}_2    & 0 \\
0         &  -\hat{\mathbb{I}}_2 \\
\end{pmatrix},
\end{equation*}
where $\hat{\mathbb{I}}_2$ is the $2\times 2$ identity matrix. It can be verified that Hamiltonian ${H}_p$ satisfies
\begin{equation}\label{Hptheta}
    \Theta {H}_p(-t)\Theta=-{H}_p(t).
\end{equation}
Consider the formal expression of the evolution operator
\begin{equation*}
    U(T,0)=\lim_{dt\rightarrow 0} \prod_{n=0}^{T/dt}e^{-i{H}_p(t_n)dt}, 
\end{equation*}
where $t_n=ndt$ and the product is time ordered. By inserting the resolution of the identity $\hat{\mathbb{I}}=\Theta\Theta$ after each factor in the product, and using relation (\ref{Hptheta}), we get
\begin{equation*}
\begin{aligned}
           \Theta U(T,0) \Theta =& \lim_{dt\rightarrow 0} \prod_{n=0}^{T/dt} \Theta e^{-i{H}_p(t_n)dt}\Theta\\
           =&\lim_{dt\rightarrow 0} \prod_{n=0}^{T/dt}  e^{i{H}_p(-t_n)dt} = U^\dag(0,-T)
\end{aligned}
\end{equation*}
Let us take the trace of the evolution operator multiplied by $\Theta$, i.e.,
\begin{equation*}
\begin{aligned}
        \text{Tr}\left[U(T,-T)\Theta\right]=& \text{Tr}\left[U(0,-T)\Theta U(T,0)\right]\\
        =& \text{Tr}\left[U(0,-T)\Theta U(T,0)\Theta\Theta\right]\\
        =&\text{Tr}\left[U(0,-T) U^\dag(0,-T)\Theta\right]=\text{Tr}\left[\Theta\right].
\end{aligned}
\end{equation*}
This implies that the $S$ matrix satisfies $\text{Tr}\left[S\Theta\right] = \text{Tr}\left[\Theta\right]=0$, which imposes a constrain on the diagonal elements,
\begin{equation}\label{one-rel3}
    S_{11}+S_{22}-S_{33}-S_{44}=0.
\end{equation}

(iv) Finally, the particular model (\ref{supp:Ham4}) has an additional symmetry. By the definition of the $S$ matrix, we have
\begin{equation*}
\begin{aligned}
       \hat{d}_p(+\infty)&=S_{31}\hat{c}_p+S_{32}\hat{c}_{-p}^{\dagger}+S_{33}\hat{d}_p+S_{34}\hat{d}_{-p}^{\dagger},\\
\hat{d}^{\dagger}_{-p}(+\infty)&=S_{41}\hat{c}_p+S_{42}\hat{c}_{-p}^{\dagger}+S_{43}\hat{d}_p+S_{44}\hat{d}_{-p}^{\dagger}. 
\end{aligned}
\end{equation*}
Taking the Hermitian conjugate of $\hat{d}_p(+\infty)$, and replacing $p$ with $-p$, we find
\begin{equation*}
    \hat{d}^{\dagger}_{-p}(+\infty)=S^*_{31}(-p)\hat{c}^{\dagger}_{-p}+S^*_{32}(-p)\hat{c}_{p}+S^*_{33}(-p)\hat{d}_{-p}^{\dagger}+S^*_{34}(-p)\hat{d}_{p},
\end{equation*}
where we used the fact that $S_{33}(-p)$ is real \cite{onepoint-LZ}. Comparing the above two different expressions for $d^\dag_{-p}(+\infty)$, we find that
\be
S_{33}(-p) = S_{44}(p).
\label{one-rel4}
\ee

Relations (\ref{be}) - (\ref{one-rel4}) are still insufficient for finding all elements of the scattering matrix. Nevertheless, they allow us to find  $\sum_p |S_{13}|^2$ without any approximation. Since the summation index in Eq.~(\ref{mag1}) in the main text runs over a symmetric interval $p\in(-\pi,\pi)$, we have

\begin{equation}
\begin{aligned}
       \sum_{p} |S_{13}|^2 \stackrel{(\ref{one-rel1})}{=}& \sum_{p}\left[e^{-\frac{2\pi |\gamma|^2}{b_1+b_2}}- S_{11}S_{33} \right]\\
\stackrel{(\ref{be2},\ref{one-rel4})}{=}&  \sum_{p>0}\left[2e^{-\frac{2\pi |\gamma|^2}{b_1+b_2}}- S_{11}(S_{33}+S_{44}) \right]\\
\stackrel{(\ref{be2},\ref{one-rel3})}{=}& 2\sum_{p>0}\left[e^{-\frac{2\pi |\gamma|^2}{b_1+b_2}}- |S_{11}|^2 \right].
\end{aligned}
\end{equation}

\section{Bounds on $N^o_{\rm ex}$}\label{app:bounds}

In this section, we discuss the bounds on $N^o_{\rm ex}$, the number of nonadiabatic excitations on the odd-site sublattice.

\subsection{Lower bound}
As given in the main text, the number of excitations on the odd-site sublattice is expressed in terms of the $S$ matrix elements,
\begin{equation}
     N^o_{\rm ex}(+\infty) = \sum_p (|S_{33}|^2 +|S_{31}|^2).
\end{equation}
The sum over the second term  in the above equation can be obtained exactly using the symmetry of the transition probability matrix (Eq.~(13) in Ref.~\cite{onepoint-LZ}): $|S_{31}|=|S_{13}|$, which is given by Eq.~(\ref{eq:s13}) in the main text.
 
The exact formula for $\sum_p|S_{33}|^2$ cannot be derived because energy level $3$ does not have the highest slope. Nevertheless, a lower bound can be found. Since $S_{33}$ and $S_{44}$ are both real,
\begin{equation}
\begin{aligned}
\sum_p |S_{33}|^2 = & \sum_{p>0}(|S_{33}|^2 + |S_{44}|^2) \\
= & \frac{1}{2}\sum_{p>0} \left[( S_{33} + S_{44})^2 + ( S_{33}-S_{44})^2 \right] \\
\ge & \frac{1}{2}\sum_{p>0} ( S_{33}+S_{44})^2=
  \sum_{p} |S_{11}|^2,
\end{aligned}
\end{equation}
where the last equality is a consequence of Eqs.~(\ref{be}) and (\ref{one-rel3}).
Thus, $N^o_{\rm ex}$ is always bounded by the number of excitations on the even-site sublattice, $ N^o_{\rm ex}\ge N^e_{\rm ex}$.

\subsection{Upper bound}
The upper bound for the number of excitations of the odd-site sublattice can be obtained under a special condition, namely, as a function of $p$, the sign of $S_{33}$ remains fixed in a given regime of the quenching rate. In this case, as a consequence of relation (\ref{one-rel4}), $S_{33}S_{44}>0$.
Hence, 
\begin{equation}
         \sum_p |S_{33}|^2 = \sum_{p>0}( |S_{33}|^2+|S_{44}|^2)
         \le  \sum_{p>0} ( S_{33}+S_{44})^2=  \sum_{p>0} |2S_{11}|^2= 2 \sum_{p} |S_{11}|^2,
\end{equation}
which is suppressed exponentially in the adiabatic limit. This condition is fulfilled for large quenching rate, for which $S_{33}\sim 1$. We thus conclude that for relatively large quenching rates, an exponential suppression for the nonadiabatic excitation on the odd-site sublattice may be observed. However, in the adiabatic limit, when the value of $S_{33}$ becomes small, this condition may be broken, which makes it possible for the power-law excitation tail to appear in the nearly adiabatic regime.

\section{Exact solutions for $J_1=J_2$}\label{app:exact}

Finally, we note that the case with $J_1=J_2$ 
can be fully solved, i.e., all elements of the $S$ matrix can be obtained analytically.
Let us group the transition probabilities into a matrix with elements $\hat{P}_{ij}\equiv |S_{ij}|^2$.
For our 4$\times$4 model at $J_1=J_2$, they are listed in Ref.~\cite{four-LZ}:
\begin{equation}
\hat{P} =
\begin{pmatrix} 
p_1p_2     & 0            & p_2q_1     & q_2  \\
0         & p_1p_2        & q_2    & p_2q_1  \\
  p_2q_1  & q_2        & p_1p_2  & 0       \\
 q_2 & p_2q_1    & 0     & p_1p_2     \\ 
\end{pmatrix},
\end{equation}
where 
\begin{equation}
    p_1 \equiv e^{-\frac{2\pi |g|^2}{|b_1-b_2|}},\ p_2 \equiv e^{-\frac{2\pi |\gamma|^2}{|b_1+b_2|}},\ q_n\equiv 1-p_n,
\end{equation}
with restriction $b_1+b_2>0$, $b_1-b_2>0$.
This solution can be extended to a more general condition with $b_1\ne\pm b_2$, by permuting the levels to full fill the condition. Now, with this exact solution, the nonadiabatic excitations at $t=+\infty$ with the initial ground state $|\downarrow\downarrow\downarrow\cdots\downarrow\downarrow\rangle$, Eqs.~(\ref{mag1}) and (\ref{n2}) in the main text, can be computed as
\begin{equation}
           N^{e(o)}_{\rm ex}(+\infty) = \sum_p e^{-\frac{2\pi |\gamma|^2}{|b_1+b_2|}}, \ b_1 \ne\pm b_2,
\end{equation}
which confirms our analytical predictions in the main text for the case with $J_1=J_2$. The above solution for the nonadiabatic excitations covers the case of $b_1=b_2$, which corresponds to the simple uniform Ising chain. 


\section{Nonadabatic excitations at $t=0$}
\label{app:kinks}

The quasi-particle excitations at $t=0$ take the form of superpositions of kinks. Here, we look at two different types of kinks separately, namely, moving from left to right along the chain, the kinks between odd and even sites, and the kinks between even and odd sites. The corresponding operators are
\begin{equation}
    \begin{aligned}
&N^e_{\rm kinks}=\frac{1}{2}\sum_{n}(1-\hat{\sigma}^x_{2n}\hat{\sigma}^x_{2n+1}),\\
&N^o_{\rm kinks}=\frac{1}{2}\sum_{n}(1-\hat{\sigma}^x_{2n-1}\hat{\sigma}^x_{2n}).
    \end{aligned}
\end{equation}
Using the Jordan-Wigner transformation (\ref{app:JWT}) and the Fourier transformation (\ref{app:FT}), the operators for the numbers of kinks can be expressed in terms of the fermionic operators, i.e.,
\begin{equation}
\begin{aligned}
&N^e_{\rm kinks}= \frac{1}{2}\sum_{p}1-(\hat{c}^\dag_p\hat{d}_p+\hat{c}^\dag_p\hat{d}^\dag_{-p}+h.c.),\\
&N^o_{\rm kinks}=\frac{1}{2}\sum_{p}1-(\hat{c}^\dag_p\hat{d}_pe^{-ip}+\hat{c}_p\hat{d}_{-p}e^{ip}+h.c.).
\end{aligned}
\end{equation}

Now the evolution of the fermionic operators in the Heisenberg picture is governed by the effective Schr\"{o}dinger equation with Hamiltonian $H_p$ in (\ref{supp:Ham4}).
Define the $S$ matrix for the evolution over the time interval $t\in(-T,0)$, i.e.,  \begin{equation}
s = s(p) \equiv  U^p (0,-T)_{T\rar \infty},
\label{s-def}
\end{equation}
with $U^p(0,-T)$ the evolution matrix with the Hamiltonian $H_p$.
The fermionic operators in the Heisenberg picture at $t=0$ admit solutions in terms of the $S$ matrix elements. We only list the relative quantities here:
\begin{equation}
\begin{aligned}
\hat{c}^\dag_p(0)\hat{d}_p(0) &= s^*_{11}s_{31}\hat{c}^\dag_p\hat{c}_p + s^*_{13}s_{33}\hat{d}^\dag_p\hat{d}_p + \hat{O},\\
\hat{c}^\dag_p(0)\hat{d}^\dag_{-p}(0) &= s^*_{11}s_{41}\hat{c}^\dag_p\hat{c}_p + s^*_{13}s_{43}\hat{d}^\dag_p\hat{d}_p + \hat{O},\\
\hat{c}_p(0)\hat{d}_{-p}(0) &= s_{12}s^*_{42}\hat{c}^\dag_{-p}\hat{c}_{-p} + s_{14}s^*_{44}\hat{d}^\dag_{-p}\hat{d}_{-p} + \hat{O}.
\end{aligned}
\end{equation}
Here, the operators on the RHS are interpreted as at $t=-\infty$ and the operator $\hat{O}$ involves all other terms that do not contribute to the computation of the number of kinks. Namely, we will evaluate the operator expectation value with respect to the initial ground state at $t=-\infty$, which is annihilated by $\hat{O}$. 
Taking average of the operators for the number of kinks with respect to the state at $t=0$, we get the number of kinks,
\begin{equation}
    \begin{aligned}
    &N^e_{\rm kinks}= \frac{1}{2}\sum_p 1-(s^*_{11}s_{31}+s^*_{13}s_{33}+s^*_{11}s_{41}+s^*_{13}s_{43}+c.g.),\\
&N^o_{\rm kinks}=\frac{1}{2}\sum_p 1-(s^*_{11}s_{31}e^{-ip}+s^*_{13}s_{33}e^{-ip}+s_{12}s^*_{42}e^{ip}+s_{14}s^*_{44}e^{ip}+c.g.)
    \end{aligned}
\end{equation}

\begin{figure}[t!]
    \centering
    \includegraphics[width=0.4\columnwidth]{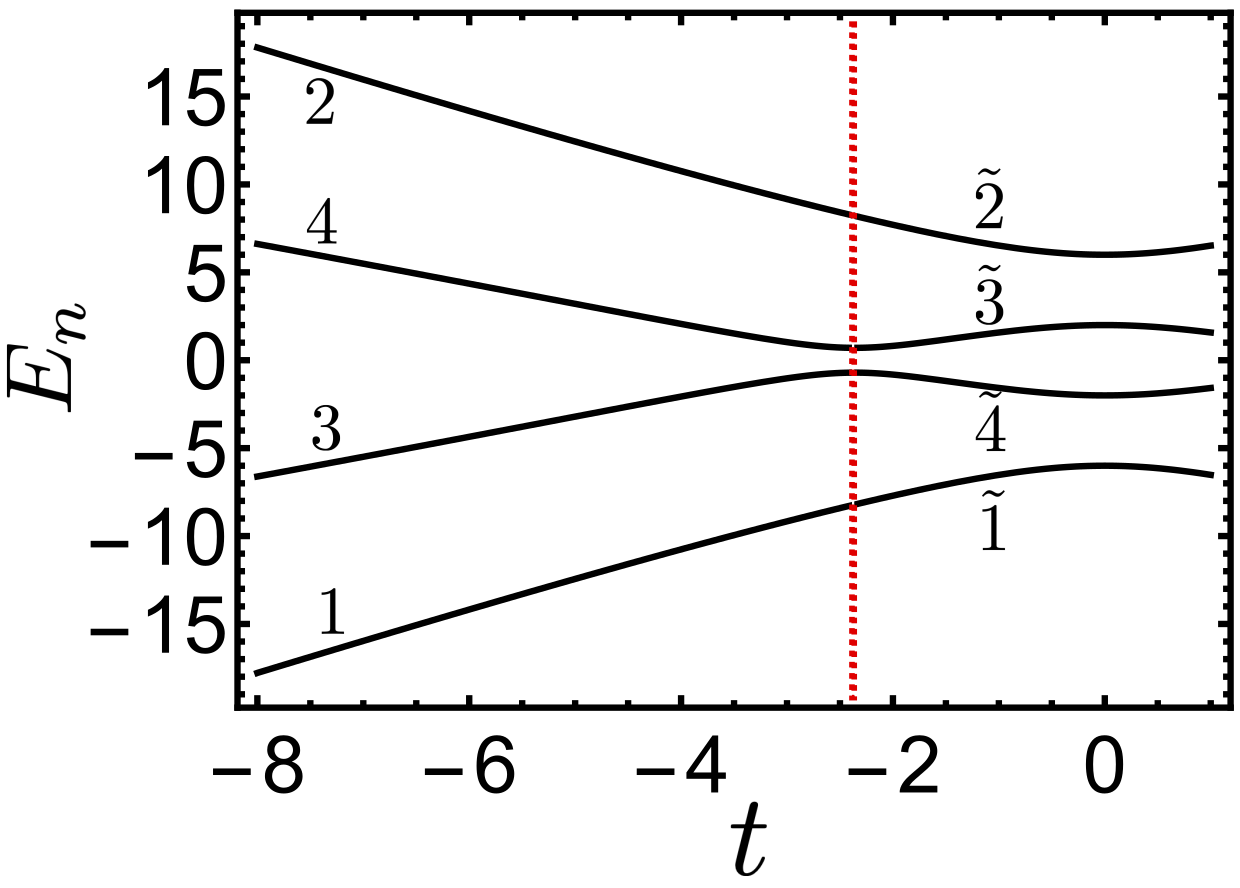}
    \caption{The time-dependent spectrum of the Hamiltonian $H_p(t)$ from~(\ref{supp:Ham4}) around the Dirac point (vertical red-dashed line), with the parameters $p=0.3$, $b_1=2$, $b_2=1$, $J_1=3$, $J_2=1$.}
    \label{fig:sup_spect}
\end{figure}

The above involved $S$ matrix elements are represent in the diabatic basis, i.e., the eigenbasis of Hamiltonian $H_p$ at $t=\infty$. One can express the $S$ matrix element in terms of transition amplitude to the eigenbasis at $t=0$, i.e., $\{|\tilde{i}\rangle\}_{i=1,2,3,4}$, as labeled in Fig.~\ref{fig:sup_spect}. For instance, 
\begin{equation}\label{eq:TranAmp}
    s_{11}=\langle 1|s|1\rangle=\langle 1|s\sum_i|\tilde{i}\rangle\langle\tilde{i}|1\rangle =\sum_i s_{1\tilde{i}}\langle\tilde{i}|1\rangle.
\end{equation}
Here the transition probability $s_{1\tilde{i}}$ can be solved using LZ theory. Finally, in terms of transition amplitudes, the number of kinks reads
\begin{equation}
\begin{aligned}
           & N^e_{\rm kinks} = \frac{1}{2}\sum_p 1- \frac{1}{4}\left[ (e^{-ip}-1)(s_{3\tilde{3}}+s_{4\tilde{3}}) - (e^{-ip}+1)(s_{3\tilde{4}}+s_{4\tilde{4}}) + c.g. \right],\\
       & N^o_{\rm kinks} = \frac{1}{2}\sum_p 1- \frac{1}{4}\left[ (e^{-ip}-1)(e^{-ip}s_{3\tilde{3}}-e^{ip}s_{4\tilde{3}}) - (e^{-ip}+1)(e^{-ip}s_{3\tilde{4}}-e^{ip}s_{4\tilde{4}}) + c.g. \right].
\end{aligned}
\end{equation}

The excitation of kinks only involves transition amplitudes between the third the fourth level around the Dirac point. This, as has been shown in the main text for excitations at $t=+\infty$ on the odd-site sublattices, results in a power-law scaling.